\documentclass[
 reprint,
superscriptaddress,
 amsmath,amssymb,
 aps,
prb,
]{revtex4-1}
\usepackage{dcolumn}
\usepackage{hyperref} 
\usepackage{graphicx}
\usepackage{amsmath}
\usepackage{siunitx}
\usepackage{bm}
\usepackage{textcomp} 
\newcommand{\eq}[1]{\begin{equation}#1\end{equation}} 
\usepackage{chemformula}
\newcommand{\chem}[1]{\ensuremath{\mathrm{#1}}}

\begin{document}


\title{Dynamic thermal relaxation in metallic films at sub-kelvin temperatures}


\author{L. B. Wang}
\email[]{libin.wang@aalto.fi}
\affiliation{QTF Centre of Excellence, Department of Applied Physics, Aalto University, FI-00076 Aalto, Finland}
\author{D. S. Golubev}
\affiliation{QTF Centre of Excellence, Department of Applied Physics, Aalto University, FI-00076 Aalto, Finland}
\author{Y. M. Galperin}
\affiliation{Department of Physics, University of Oslo, PO Box 1048, Blindern, 0316 Oslo, Norway}
\affiliation{A. F. Ioffe Physical-Technical Institute, Russian Academy of Science, St. Petersburg 194021, Russia}
\author{J. P. Pekola}
\affiliation{QTF Centre of Excellence, Department of Applied Physics, Aalto University, FI-00076 Aalto, Finland}

\date{\today}

\begin{abstract}
The performance of low temperature detectors utilizing thermal effects is determined by their energy relaxation properties. Usually, heat transport experiments in mesoscopic structures are carried out in the steady-state, where temperature gradients do not change in time. Here, we present an experimental study of dynamic thermal relaxation in a mesoscopic system -- thin metallic film. We find that the thermal relaxation of hot electrons in copper and silver films is characterized by several time constants, and that the annealing of the films changes them. In most cases, two time constants are observed, and we can model the system by introducing an additional thermal reservoir coupled to the film electrons. We determine the specific heat of this reservoir and its coupling to the electrons. The experiments point at the importance of grain structure on the thermal relaxation of electrons in metallic films.
\end{abstract}


\maketitle
\section{Introduction}
Investigation of thermal transport in mesoscopic devices is essential for the study of fundamental physics and for the development of low temperature detectors \cite{Giazotto2006}. Examples include transition edge sensors \cite{Irwin2005,Karasik2011,Ullom2015,Miaja-Avila2016,Morgan2018}, hot-electron detectors \cite{Nahum1993,Kuzmin1999,Govenius2016,Efetov2018,Kokkoniemi2019} and other types of nano-calorimeters in the application of photon and particle detection \cite{Karasik2012,Doriese2017,Kuzmin2019}, mass spectrometry \cite{Hilton1998,Novotny2015} and in the field of quantum technologies \cite{Narla2016,Romero2009,Barzanjeh2015,Opremcak2018}. In the steady-state, thermal transport down to a single heat-conducting channel has been studied at low temperatures in the last decades \cite{Schwab2000,Chiatti2006,Jezouin2013,Mosso2017,Banerjee2017}. In the dynamic regime, tremendous progress has been achieved in investigating electron-phonon interactions in metals at picosecond time scales with laser pumping techniques \cite{Rosei1972,Eesley1986,Schoenlein1987,Elsayed-Ali1987,Allen1987,Gamaly2011}. In this paper we focus, instead, on milli- and microsecond time scales and at very low temperatures. Time resolved experiments in this regime became possible with the recent development of fast low temperature thermometers \cite{Gasparinetti2015,Schmidt2003,Wang2018,Zgirski2018,Karimi2018}, and they are particularly important for further improvement of low temperature detectors utilizing thermal effects. Indeed, thermal relaxation determines the performance of the detector and its recovery time after a detection event. In the absence of detailed time resolved measurements, the relaxation in the detector is usually assumed to be exponential, and its time constant is estimated from the parameters measured in the steady state \cite{Schmidt2003,Wei2008}.
Here we demonstrate the deficiency of this approach for a metallic film absorber, 
which may exhibit several relaxation times and may equilibrate very slowly. 

\begin{figure}[t]
\includegraphics[width= \columnwidth]{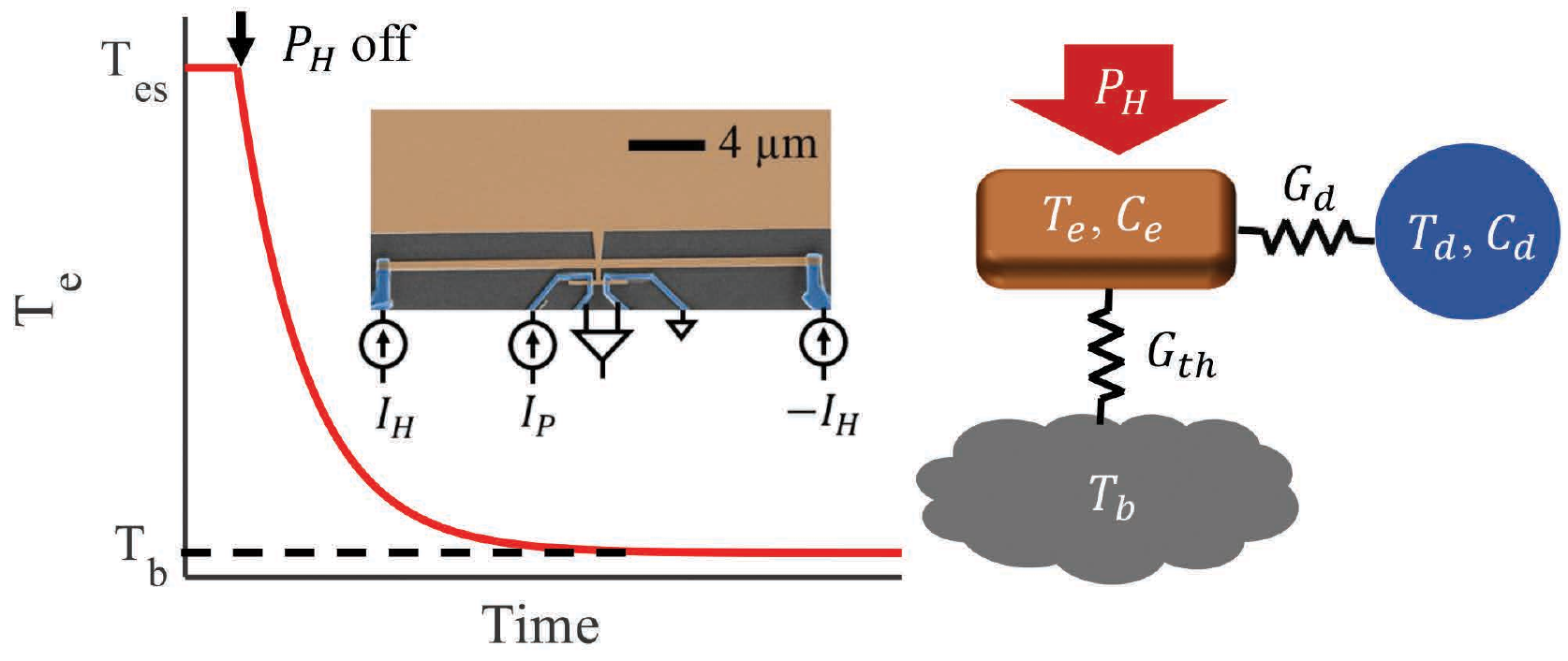}
\caption{Left panel:  Time dependence of the electron temperature $T_e$ in response to the external heating power $P_H$. After $P_H$ is switched off, the electron temperature relaxes to the bath temperature $T_b$. Inset: False-color SEM image of a typical sample with the measurement circuits. (Cu is shown in brown, Al in blue). Right panel: Thermal model of a heated Cu film on a dielectric substrate. Film electrons are thermally coupled to the bath ($T_b$) and to an additional thermal reservoir ($T_d, C_d$) by thermal conductances $G_{th}$ and $G_d$.}
	\label{Fig1}
\end{figure}

In this article, we mostly focus on the thermal relaxation process in a copper (Cu) film and study the time dependence of the electron temperature after the applied heating is turned off, as shown in Fig.~\ref{Fig1}. Theory predicts that for a normal metal at low temperatures, thermal relaxation is governed by electron-phonon (e-ph) coupling \cite{Wellstood1994}, then the relaxation time is given by $\tau_{e-ph} = \gamma/5\Sigma T^3$. Here $\gamma$ is the Sommerfeld constant, which determines the heat capacity of electrons,  $C_e=\gamma VT$, with $V$ the volume of the film, and $\Sigma$ the characteristic electron-phonon coupling constant governing the heat flux $P_{e-ph}$ between electron and phonon sub-systems having the temperatures $T_e$ and $T_{ph}$,  $P_{e-ph}=\Sigma V(T_e^5-T_{ph}^5)$. For Cu, these two parameters take the values $\gamma$ = 98 JK$^{-2}$m$^{-3}$ \cite{Kittel2005} and $\Sigma$ = 2~nWK$^{-5}$$\mu$m$^{-3}$ \cite{Wellstood1994,Roukes1985}, which results in a relaxation time $\tau_{e-ph} \sim 10$ $\mu$s at 100 mK. In our experiment, we show, contrary to the theoretical prediction, that the relaxation of electron temperature in Cu films is characterized by several time constants, and annealing the film will change the relaxation dramatically. We have observed the same effects in silver (Ag) films with lattice dislocations. Our experimental results indicate that, at low temperatures, the thermal relaxation of electrons in a metallic film is strongly affected by the film grain structure.

\section{Experiments with Copper films}
One of our devices is shown as a false-color SEM image in the inset of Fig.~\ref{Fig1}. Cu films of varying thicknesses are evaporated on \chem{SiO_2/Si} substrate with the electron-beam evaporation technique. Superconducting aluminum (Al) is used for the galvanic connection to the Cu films. The large pad on top (partly shown) constitutes the main volume of the film to be studied. Long horizontal Cu wire in the middle is a heater, it is biased with currents of opposite polarities $\pm I_{H}$. The short wire at the bottom, which contacts two aluminum leads at both ends, behaves as a proximity Josephson junction (JJ). It is used as a fast thermometer to monitor electron temperature in the Cu film. In the explored temperature range its resolution was around 0.1~mK for a probing pulse of 2~$\mu$s duration, shown in Appendix~\ref{appendix:a}. Details of the measurement technique have been reported earlier \cite{Wang2018}. Thermal equilibration times between the heater, the thermometer and the pad can be estimated as $\tau_{eq}^{ij} = C_i/G_{ij}$, where $C_i$ is the heat capacity of the corresponding film and $G_{ij}$ is the heat conductance of the bridge connecting the two components. We find $\tau_{eq}^{ij}<300$ ns at 0.1~K for all measured devices (see discussion in Appendix~\ref{appendix:b}), which is much shorter than the thermal relaxation times of interest in this work.

\begin{figure}[h]
	\centering
	\includegraphics[width= \columnwidth]{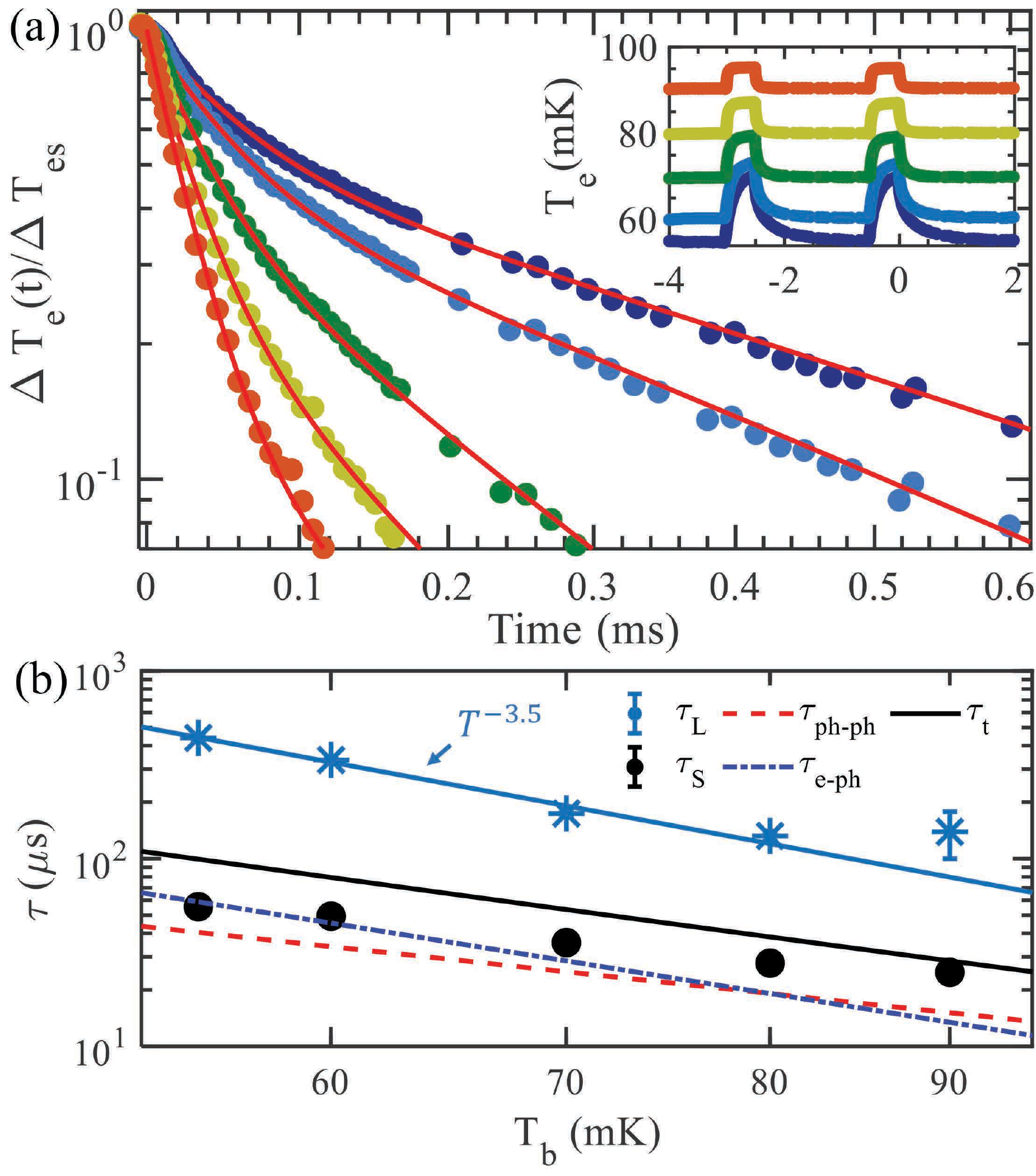}
	\caption{Thermal relaxation in a 300~nm thick Cu film. (a) Time dependence of the normalized electron temperature $\Delta T_e(t)/\Delta T_{es}$. Bath temperatures are $T_b$ = 55, 60, 70, 80, 90~mK from right to left. Red lines are fits with Eq. (\ref{eq1}). Inset: electron temperature versus time, the values of $T_b$ are the same as in the main plot. (b) Temperature dependence of the two time constants $\tau_L$ and $\tau_S$. $\tau_{e-ph}$ (dash-dotted), $\tau_{ph}$ (dashed) and $\tau_t$ (solid black) are the predicted time constants given by Eqs. (\ref{tau_t}).  The solid-blue line shows a simple power law fit of the dependence $\tau_{L}(T)$.} \label{Fig2}
\end{figure}

In the inset of Fig.~\ref{Fig2} (a) we show the time dependence of the electron temperature to the two heating current pulses with inverse polarities. Each current pulse had a width of $500~\mu$s, and it was applied during the time intervals $-3.5$ to $-3$~ms and$ -0.5$ to $0$~ms. The thickness of the Cu film was 300~nm. Before the heating pulse was applied, i.e., for $-4 < t < -3.5$~ms, the electrons were in thermal equilibrium having the bath temperature $T_b$. After the heating was turned on, the electron temperature began to rise and finally reached the steady-state value $T_{es}$. Identical responses to the two heating pulses confirm the accuracy of our measurement technique. The steady-state temperature rise $\Delta T_{es} = T_{es} - T_b$ is consistent with the previous DC measurements \cite{Wang2019}. It decreases at high $T_b$ due to the increase of the thermal conductance between the electrons in the film and the environment.

To investigate the dynamic thermal relaxation, we have recorded the time dependence of the electron temperature after the heating current is turned off.  Figure~\ref{Fig2} (a) shows the normalized $\Delta T_e(t)$ as a function of time $t$. We have found, contrary to our expectation, that the dependence $\Delta T_e(t)$ could not be fitted with a single exponentially decaying function. However, we could very well fit the data with two exponentials,
\eq{
\Delta T_e (t)/\Delta T_{es}= ae^{-\frac{t}{\tau_L}} + (1-a)e^{-\frac{t}{\tau_S}}, \label{eq1}
}
where $a$ is a constant pre-factor, and $\tau_S$ and $\tau_L$ are, respectively, the short and the long relaxation times. Red lines in Fig.~\ref{Fig2} (a) show the corresponding fits. The times $\tau_S$ and $\tau_L$ for different temperatures are shown in Fig.~\ref{Fig2} (b). At low temperatures, the time constant $\tau_L$ is about one order of magnitude longer than $\tau_S$, and at high temperatures $\tau_L$ and $\tau_S$ differ even more. As a result, at high temperatures $\tau_S$ cannot be resolved and the relaxation process takes a single exponential form with a time constant $\tau_L$. This may explain the previous observations of very long relaxation times in Cu and AuPd films \cite{Gasparinetti2015,Govenius2016}. Due to the high sensitivity of our JJ thermometer at low temperatures, we can now clearly distinguish the two time constants. Performing simple power law fits, we find that the relaxation times scale with temperature as $\tau_L \propto T^{-3.5}$ and  $\tau_S\propto T^{-3}$. We have also verified that the effect of the heating pulse amplitude on thermal relaxation is negligible for temperature increments within the range 6 mK $<\Delta T_{es}<$ 18 mK, as discussed in Appendix~\ref{appendix:c}.
\begin{figure}[t]
	\centering
	\includegraphics[width= \columnwidth]{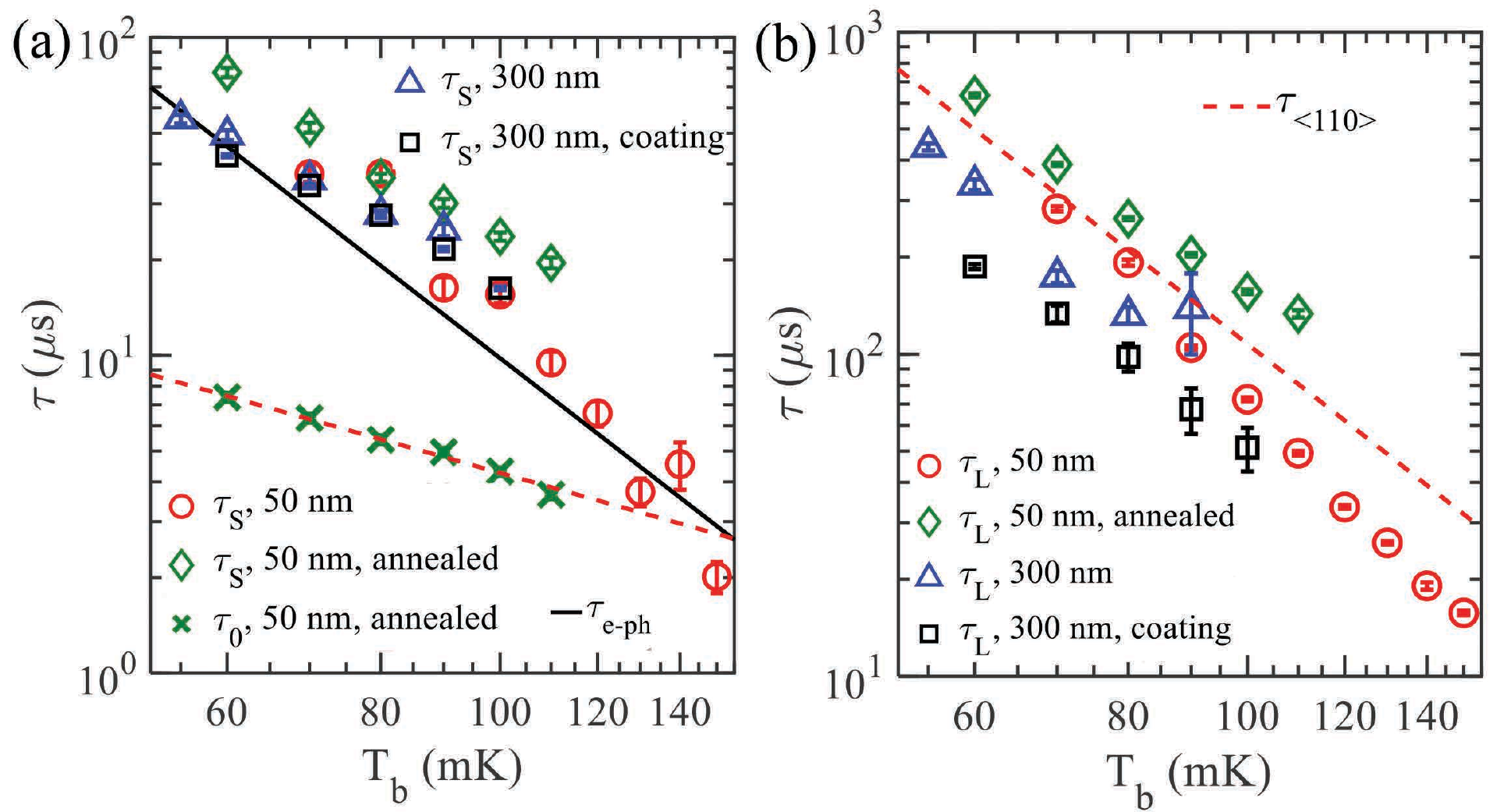}
	\caption{Temperature dependence of thermal relaxation times for all samples. (a) Short time constant $\tau_S$ (for the annealed 50~nm film we show both $\tau_S$ and the shortest relaxation time $\tau_0$). Black-solid line shows the calculated electron-phonon time $\tau_{e-ph}$ (\ref{tau_t});  red-dashed line is the linear fit of $\tau_0$ of the annealed 50~nm film. (b) Longer realxation time $\tau_L$ versus temperature. Red-dashed line shows the electron-phonon relaxation time of electrons moving in $\langle110\rangle$ direction measured in bulk Cu, $\tau_{\langle 110\rangle} = 10^{-7} \times (T/1{\rm K})^{-3}$ s \cite{Doezema1972}.}
	\label{Fig3}
\end{figure}

We have repeated the measurements with several films in order to test
the dependence of the relaxation times on the film thickness, which has been observed earlier \cite{Viisanen2018}. Such dependence would
suggest that multiple relaxation times may be caused by impurities or defects on the surface of the film. To verify this, we have fabricated a sample with a film thickness of 50~nm, in which the surface to volume ratio is thus increased by a factor of 6, and a sample with 300~nm Cu film coated with a 5~nm thin layer of Al right after evaporation. The volumes of these two films are the same as the volume of the original 300~nm film, i.e. 120 $\mu$m$^3$.
We have found that both films show two time constants, which are quite close to the ones of the 300 nm film without surface coating, as shown in Fig.~\ref{Fig3} (a).
Finally, we have measured an annealed 50 nm thin Cu film in order to test the effect of the grain size and grain interface on thermal relaxation. Appendix~\ref{appendix:d} shows the SEM images of the sample with significant growth of the grain size after annealing. Thermal relaxation of the annealed film has changed significantly, and we needed three exponentials with different relaxation times in order to fit the dependence of $\Delta T_e(t)$ with sufficient accuracy. These time constants are also shown in Fig.~\ref{Fig3}. The longest and the middle of them are of the same order of magnitude as the times $\tau_L$ and $\tau_S$ measured in other samples. The shortest time constant of the annealed film $\tau_0$ is about 6 - 10 times smaller than $\tau_S$, and has a linear dependence on inverse temperature.

Summarizing these observations, we conclude that the appearance of an additional relaxation time $\tau_L$ cannot be explained by surface effects. One may alternatively relate it to magnetic impurities in the bulk of the films. However, the previous experiment \cite{Viisanen2018} has ruled out such a possibility for Cu films. It was found in Ref. \cite{Viisanen2018} that Cu films with a very low concentration of magnetic impurities still exhibited long relaxation time. In contrast, less pure silver (Ag) film had only shown short relaxation time $\tau_S$, and its value was in good agreement with the predictions of the free electron model. We argue below that the origin of the multi-scale thermal relaxation could be rather explained by the morphology of the films.

\section{Phenomenological thermal model and discussion}
In order to analyze the two scale relaxation of the electron temperature on the quantitative level, we propose a phenomenological model, in which the electrons in the Cu film are coupled to phonons and to an additional thermal reservoir, as shown in Fig.~\ref{Fig1}. In this model, the time evolution of temperatures follows the equations
\begin{eqnarray}
C_e\frac{dT_e}{dt} &=& -\Sigma V(T_e^5 - T_{ph}^5) - \Sigma_{d}(T_e^\alpha - T_{d}^\alpha) + P_H(t),
\nonumber\\
C_{ph}\frac{dT_{ph}}{dt} &=& \Sigma V(T_e^5 - T_{ph}^5) - \kappa A(T_{ph}^4 - T_b^4),
\nonumber\\
C_{d}\frac{dT_{d}}{dt} &=& \Sigma_{d}(T_e^\alpha - T_{d}^\alpha).
\label{relaxation}
\end{eqnarray}
Here $C_d$ is the heat capacity of the thermal reservoir, $T_d$ is its temperature, $\Sigma_d$ is the constant characterizing the coupling between electrons and the reservoir, $A$ is the contact area between the film and the dielectric substrate, $\kappa$ is the constant characterizing thermal boundary conductance between the phonons in the film and the substrate, $C_{ph}$ is the heat capacity of the phonons in the film, $C_e=\gamma VT_e$ is the heat capacity of electrons, $\alpha$ is an unknown exponent, and $P_H(t)$ is the heating power. The phonon heat capacity is usually very small, and one can put $C_{ph}=0$. Adopting this approximation and
considering linearized versions of Eqs. (\ref{relaxation}), which are valid at sufficiently small $P_H$, we obtain $\Delta T_e(t)$ after an abrupt removal of the heating in the form of Eq.~(\ref{eq1}) with the relaxation times and the pre-factor $a$ having the form
\begin{eqnarray}
\frac{1}{\tau_{L,S}} &=& \frac{1}{2}\left[\frac{1}{\tau_d}+\frac{1}{\tau_{t}}\right] \mp \sqrt{\frac{1}{4}\left[\frac{1}{\tau_{t}} - \frac{1}{\tau_d}\right]^2 + \frac{C_d}{C_e + C_d}\frac{1}{\tau_{t}\tau_d}},
\nonumber\\
a &=& \frac{\tau_d^{-1}-\tau_L^{-1}}{\tau_S^{-1}-\tau_L^{-1}} \frac{\tau_d}{\tau_S}.
\label{eq3}
\end{eqnarray}
The time $\tau_d$, appearing above, characterizes the relaxation between electrons and the additional thermal reservoir, $\tau_d^{-1}=(C_d^{-1}+C_e^{-1})G_d$, and $\tau_t$ is the relaxation time of the electron temperature in the absence of the reservoir. It is given by the sum of two contributions,
\begin{eqnarray}
&&\tau_t = \tau_{e-ph}+\tau_{ph-ph},
\nonumber\\ &&
\tau_{e-ph} = {C_e}/{G_{e-ph}},\;\; \tau_{ph-ph}= {C_e}/{G_{ph-ph}}.
\label{tau_t}
\end{eqnarray}
In the above expressions, $G_d = \alpha\Sigma_d T^{\alpha-1}$ is the thermal conductance between electrons and the reservoir, 
$G_{e-ph}=5\Sigma VT^4$ is the thermal conductance between electrons and phonons, and $G_{ph-ph}=4\kappa AT^3$ is the thermal conductance
between the phonons in the film and the substrate. Note that if one keeps the finite value of the phonon heat capacity $C_{ph}$,
then an additional time constant (\ref{eqE4}) formally appears in the model. However, this time constant is very short and
cannot be identified with any of the relaxation times $\tau_S$ or $\tau_L$, see Appendix~\ref{appendix:e} for detailed discussion.
In Fig.~\ref{Fig2} (b) we plot the times $\tau_{e-ph}$, $\tau_{ph-ph}$ and $\tau_t$, given by Eqs. (\ref{tau_t}), with the constants $\Sigma$ = 2~nWK$^{-5}\mu$m$^{-3}$ and $\kappa$ = 60~pWK$^{-4}\mu$m$^{-2}$ measured in an independent steady-state experiment \cite{Wang2019}.  We observe rather good agreement between the measured values of $\tau_S$ and the calculated times $\tau_t$.

\begin{figure}[t]
	\centering
	\includegraphics[width= \columnwidth]{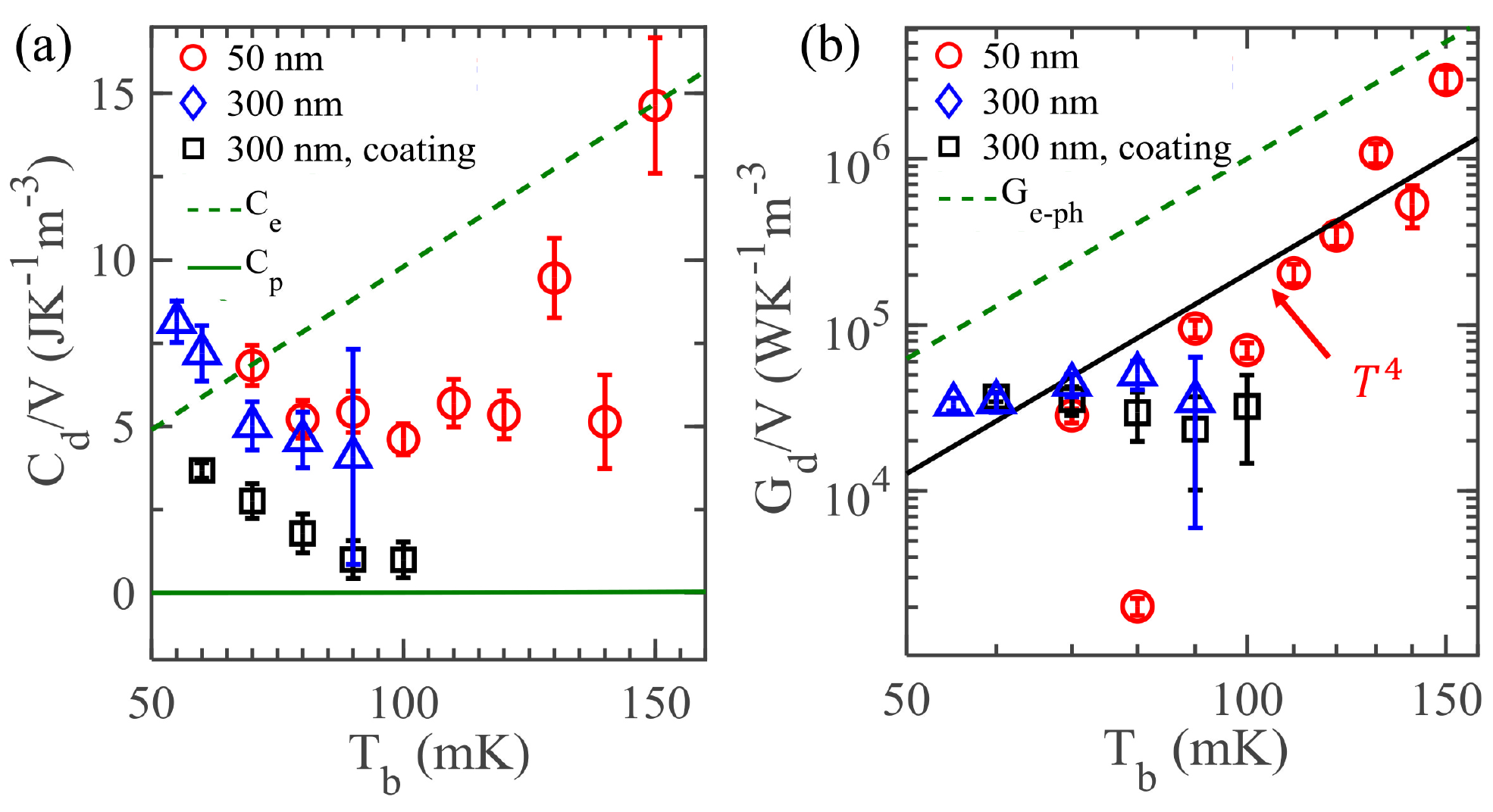}
	\caption{(a) Temperature dependence of the specific heat of the additional thermal reservoir coupled to electrons (\ref{Cd}), i.e., heat capacity normalized by the volume of the absorber, $C_d/V$, for three different Cu films. Solid and dashed lines are the phonon (see Eq. \ref{Cph}) and electron specific heats of the films, respectively. (b) Thermal conductance between the reservoir and the electrons  (\ref{Gd}) per unit volume, $G_d/V$. The dashed line shows the electron-phonon thermal conductance $G_{e-ph}/V$; the solid line is the fit of the data for the 50~nm thin film with $\propto T^4$ dependence.}\label{Fig4}
\end{figure}

Next, we invert Eqs.~(\ref{eq3},\ref{tau_t}) and express the heat capacity of the thermal reservoir, $C_d$, and the thermal conductance, $G_d$,
in terms of measured parameters, 
\begin{eqnarray}
C_d &=& [(\tau_L-\tau_S)^2/\tau_L\tau_S]a(1-a) \gamma VT,
\label{Cd}
\\
G_d &=& [C_dC_e/(C_d+C_e)]\left(\tau_L^{-1}+\tau_S^{-1}-\tau_{t}^{-1}\right).
\label{Gd}
\end{eqnarray}
Using the measured data, we calculate $C_d$ and $G_d$ in order to get more information about the additional thermal reservoir. 
These values are plotted for three different films in Figs.~\ref{Fig4}~(a) and (b).
The heat capacity $C_d$ does not exhibit clear temperature dependence in the explored temperature range. Its value is rather high, it significantly exceeds the phonon heat capacity (\ref{Cph}) and turns out to be comparable to that of electrons $C_e$. It is unlikely that low concentration impurities would result in such a high value. We speculate that the most natural reason for that would be the existence of a significant number of grains in the film, which are weakly thermally coupled to each other due to electron scattering at the grain boundary. Such grains would form an additional thermal reservoir, which we have postulated in our model. In Fig. \ref{Fig4} (b) we show the temperature dependence of the thermal conductance $G_d$. For the two 300~nm  films  $G_d$ is almost constant, while for the 50~nm film it roughly follows a power-law dependence with the exponent close to 4,
which corresponds to $\alpha=5$ in Eq. (\ref{relaxation}).  

Our hypothesis about weakly coupled grains in the film as a reason for the long thermal relaxation time is supported by other experiments.
It has been shown that for evaporated Cu films on \chem{SiO_2/Si} substrate \cite{Wei2002,Huang2003}, at the beginning of the growth process, a film with $\langle111\rangle$ orientation is formed. Subsequently, grains with $\langle110\rangle$ orientation are nucleated at the boundaries of the $\langle111\rangle$ grains, which results in the growth of $\langle110\rangle$ texture. Afterward, $\langle111\rangle$ grains can again nucleate at the boundaries between $\langle110\rangle$ grains and so on. In the end, one obtains a sandwiched structure of alternating textures, which might be, sometimes, poorly thermally coupled to each other. It is also well known that electron-phonon scattering rates in bulk Cu are strongly anisotropic \cite{Doezema1972}. Since in thin films phonon wave vectors are parallel to the surface and since the Fermi velocity of electrons is much higher than the speed of sound, phonons predominantly interact with electrons moving perpendicular to the film surface. Thus, in grains with $\langle110\rangle$ orientation, for example, the scattering rate should be close to that of electrons moving in $\langle110\rangle$ direction in the bulk material. Bulk relaxation time in the $\langle110\rangle$ direction is, indeed, significantly longer than that in $\langle111\rangle$ direction, as magnetic resonance experiments have shown \cite{Doezema1972}. In Fig.~\ref{Fig3} (b), we plot measured electron-phonon relaxation time in the bulk Cu in $\langle110\rangle$ direction, $\tau_{\langle 110\rangle} = 10^{-7} \times (T/1{\rm K})^{-3}$~s \cite{Koch1970,Doezema1972,Gantmakher1973}, and find that its value is indeed close to measured $\tau_L$. Within this scenario, the high value of the pre-factor in front of slowly decaying exponent observed for the 50~nm thin Cu film, $a\approx 0.8$, points to the dominant $\langle110\rangle$ texture. For the two 300~nm films $a$ drops from $0.5$ at low $T$ to $0$ at higher $T$ (shown in Appendix~\ref{appendix:f}), which hints to $\langle111\rangle$ as preferred orientation.

Though the above arguments qualitatively explain our findings, further experiments and more detailed theoretical modeling are required to fully understand heat relaxation mechanisms in thin Cu films. For example, thermal relaxation in the annealed 50 nm film can only be fitted with three exponentials and, therefore, cannot be described by the model (\ref{relaxation}). On the other hand, the pronounced effect of annealing on thermal relaxation points at the importance of the grain structure of the film.  Yet another unclear issue is the nature of the thermal coupling between the weakly coupled grains, which is described by the thermal conductance $G_d$. In the case of coupling through tunnel barriers between the grains, one would expect $G_d\propto T$, but the observed $G_d\propto T^4$ scaling differs from that. We have also considered an alternative thermal model, in which the additional reservoir couples to film phonons instead of electrons. However, this model results in much smaller values of the pre-factor $a$  than the observed ones.

\section{Experiment with Silver film}

\begin{figure}[t]
\includegraphics[width=\columnwidth]{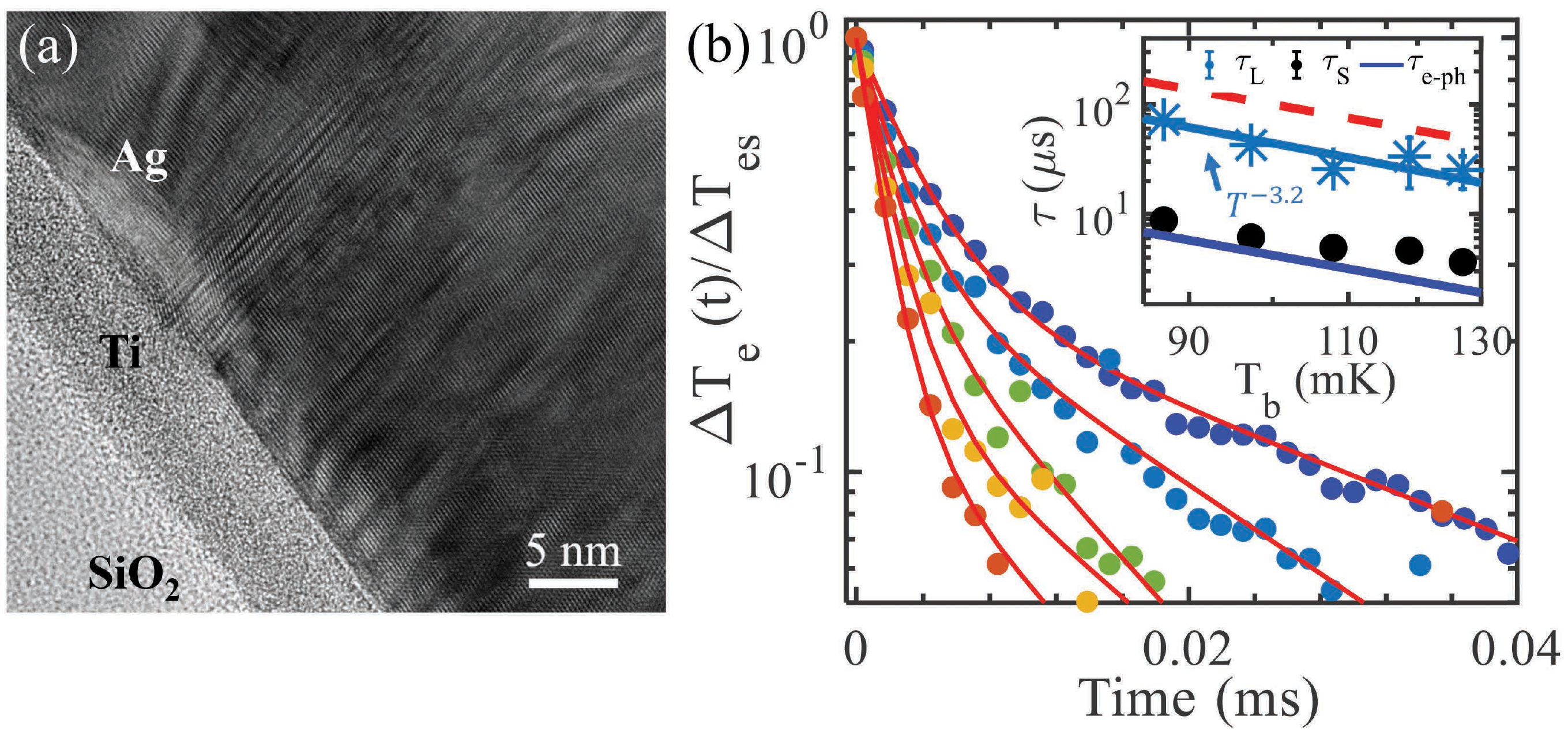}
\caption{(a) TEM image shows the Ag film with lattice dislocations. The Ag film is evaporated on a \chem{SiO_2}/Si substrate with 5~nm of Ti in-between. (b) Normalized electron temperature $\Delta T_{e}/\Delta T_{es}$ as a function of time. $T_b$ = 87, 97, 108, 119, 127~mK from right to left. Red lines are fits with Eq.~\ref{eq1}. Inset: temperature dependence of $\tau_S$ (dots) and $\tau_L$ (stars) derived from the fits.  $\tau_{e-ph} = \gamma/5\Sigma T^3$ is the calculated e-ph thermal relaxation time (dark blue solid line). For Ag, we used $\Sigma_{Ag}$ = 3~nWK$^{-5}\mu$m$^{-3}$\cite{Viisanen2018} and $\gamma_{Ag}$ = 63 JK$^{-2}$m$^{-3}$ \cite{Kittel2005}. Red-dashed line shows the e-ph relaxation time of electrons moving in $\langle110\rangle$ direction, $\tau_{<110>}$ = 1.7$\times10^{-7}$ (T/1K)$^{-3}$~s \cite{Gasparov1975}.}\label{Fig5}
\end{figure}

The importance of film grain structure on thermal relaxation is further supported by experiments with Ag film. Previous experiments \cite{Viisanen2018} have shown that in the uniform Ag film, thermal relaxation exhibited a single relaxation time. We have carried out an additional experiment with a 50~nm silver film having a granular structure. Figure~\ref{Fig5} (a) shows the transmission electron microscopy (TEM) image of that film. In contrast to the previous experiment \cite{Viisanen2018}, the film image clearly shows lattice dislocations. We have found that the electron temperature relaxation in the granular Ag film is a double exponential decay similar to that in copper films. It is illustrated in Fig.~\ref{Fig5} (b). In the same way as for the Cu films discussed, the value of the shorter relaxation time $\tau_S$ is close to the calculated e-ph thermal relaxation time $\tau_{e-ph}$, while the longer relaxation time $\tau_L$ is about one order of magnitude larger. We believe that the difference between the two times can also be explained in the same way.
Indeed, magnetic resonance experiments \cite{Gasparov1975,Johnson1976} have demonstrated strong anisotropy of the e-ph scattering in silver. In the inset of Fig.~\ref{Fig5} (b) we show the e-ph relaxation time measured in thick silver films with $\langle110\rangle$ orientation averaged over magnetic field direction, 
$\tau_{\langle 110 \rangle}\approx 10^7 (T/1{\rm K})^3$ Hz (note that this value coincides with the one measured in copper), and find that it is close to the measured $\tau_L$. Thus, our measurement of the silver film further emphasizes the importance of the grain texture of the film for thermal relaxation.

\section{Conclusion}

In summary, we have investigated dynamic thermal relaxation in Cu and Ag films at sub-kelvin temperatures. In contrast to uniform Ag thin films, which are well described by the free electron model \cite{Pinsolle2016,Viisanen2018}, thermal relaxation in granular Cu and Ag films is complicated and characterized by two or several relaxation times. Our experiment points at the importance of film grain structure on the thermal relaxation of electrons in a metallic film. Further research is needed in order to fully understand the microscopic mechanism behind these observations.
The experiment refines the understanding of non-equilibrium thermal transport in mesoscopic metallic structures and should help to further optimize the performance of devices utilizing thermal effects and operating at low temperatures.

\begin{acknowledgments}
We thank O.-P. Saira for technical help, as well as K. L. Viisanen and C. Enss for useful discussions. We acknowledge the provision of the fabrication facilities by Otaniemi research infrastructure for Micro and Nanotechnologies (OtaNano). This work was performed as part of the Academy of Finland Centre of Excellence program (Projects No.312057.) and European Research Council (ERC) under the European Union's Horizon 2020 research and innovation program (grant agreement No. 742559).
\end{acknowledgments}
%
\appendix
\section{Characterization of Josephson junction thermometer}
\label{appendix:a}
\begin{figure}[h]
\includegraphics[width=\columnwidth]{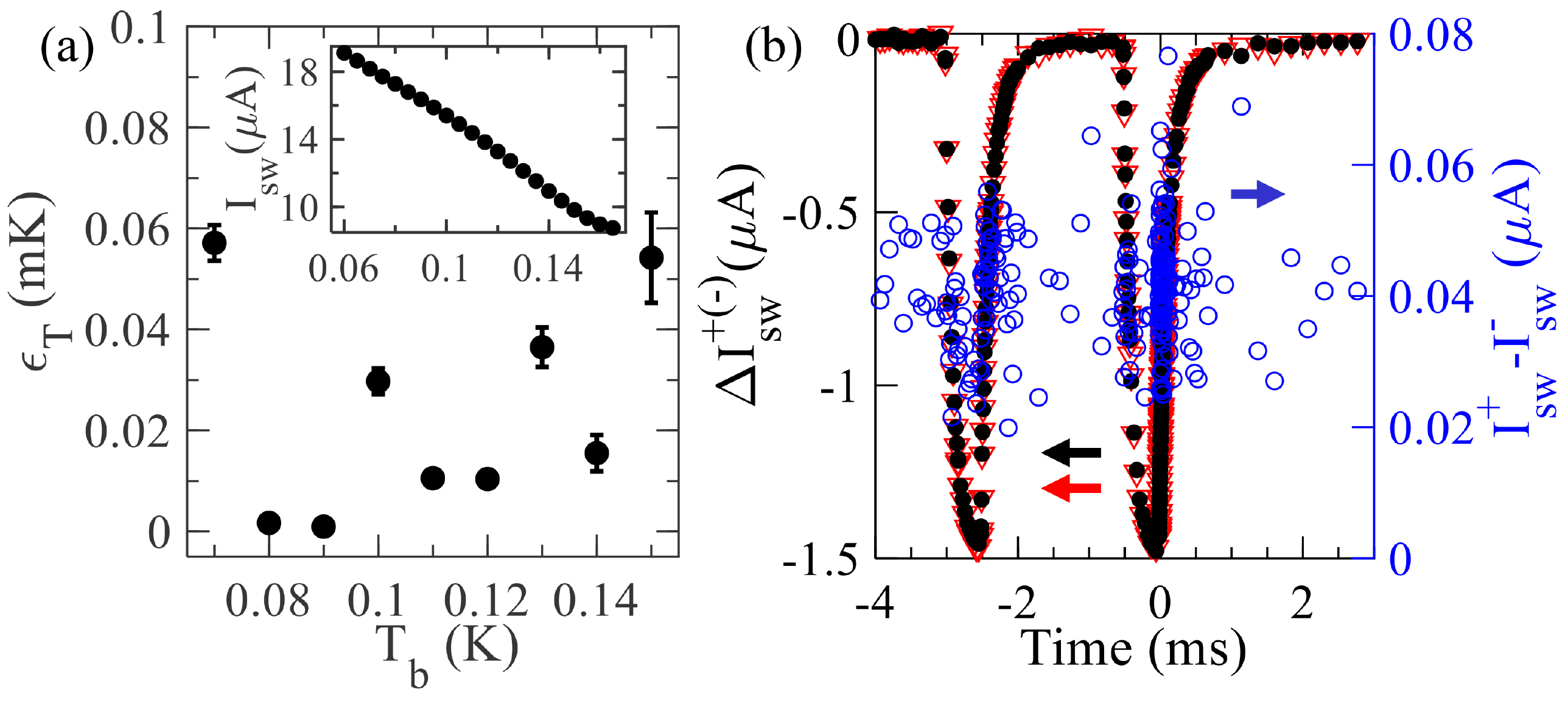}
\caption{Characterization of the PJJ thermometer. Inset of (a) shows $I_{sw}$ as a function of bath temperature $T_b$, which is used as the temperature calibration of the PJJ thermometer. (a) Temperature resolution $\epsilon_T$ of the PJJ thermometer in the temperature range explored. (b) Switching current changes $\Delta I_{sw}^{+(-)}$ and the offset as a function of the time interval between $I_H$ and $I_p$.}\label{Fig6}
\end{figure}

The switching from the superconducting to the resistive state of a proximity Josephson junction (PJJ) occurs stochastically due to both thermal and quantum fluctuations. In order to define the switching current $I_{sw}$ we have calibrated the thermometer. We have probed the switching by sending a pulse train of $N$ pulses with the fixed probing current $I_p$, and recorded the number of switchings, $n$, by monitoring the voltage across the PJJ. The switching probability at a fixed bias current is defined as $P = n/N$. 
We define the switching current, $I_{sw}$, as the bias current at which $P = 50\%$. Temperature dependence of $I_{sw}$ is shown in the inset of Fig.~\ref{Fig6} (a). No saturation of $I_{sw}$ is observed down to  60~mK. 
To characterize the performance of the PJJ thermometer we define the temperature resolution of the thermometer, 
\eq{
	\epsilon_T = \frac{\Delta I_p}{dI_{sw}/dT}, \label{eqA1}
}
where $\Delta I_p = I_p(P = 90\%) - I_p(P = 10\%)$ is the width of the switching distribution. For our underdamped PJJ, $\Delta I_p$ is around 20~nA. The calculated temperature resolution of the thermometer is shown in Fig.~\ref{Fig6} (a). In the temperature range explored, $\epsilon_T$ is around 0.1~mK for a probing pulse with a width of 2~$\mu$s. We have used heating signals, $I_H(t)$, consisting of two pulses (pulse width 500~$\mu$s) of opposite polarities to heat film electrons. Fig.~\ref{Fig6} (b) shows the measured switching current changes $\Delta I_{sw}^{+(-)} = I_{sw}^{+(-)} - I_{sw}^{+(-)}(t = 0)$ for both polarities. Here, +(-) indicates the direction of the probing current. Overlapping of $\Delta I_{sw}^{+(-)}$ confirms the accuracy of the measurements. In addition, we have calculated the switching current offset $I_{sw}^{+} - I_{sw}^{-}$ and plotted it in Fig.~\ref{Fig6} (b) with the blue circles. The offset fluctuates around an averaged value and does not depend on whether $I_H$ is turned on or off, thus indicating that the heating current flowing to the temperature probing line is negligible.

\section{Time of thermal equilibration between the heater, the thermometer and the pad}
\label{appendix:b}
The thermal equilibration time between the heater, the thermometer and the pad is estimated as $\tau_{eq}^{ij} = C_i/G_{ij}$, where $C_i$ is the heat capacity and $G_{ij}$ is the heat conductance of the bridge between the heater, the thermometer and the large pad. For metallic films at low temperatures, $C_i$ is dominated by electronic heat capacity and is estimated from the free electron model as $C_i = \gamma VT$, where $V$ is the volume of the large pad. Heat conductance $G_{ij}$ is dominated by the electrons and one can estimate it using the Wiedemann-Franz law as $G_{ij} =LTG_{el}$, where $G_{el}$ is the electronic conductance and $L = 2.44 \times10^{-8}$~W$\Omega$K$^{-2}$ is the Lorenz number. The resistivity of the narrow Cu wire is measured from the heater resistance of about $\rho$ = 1$\times$10$^{-8}~\Omega$m. For a typical bridge between the heater, the thermometer and the pad with dimensions of 700~nm $\times$ 250~nm $\times$ 50~nm, $G_{el} \approx 2$~S. At temperature of 0.1~K, the estimated $\tau_{eq}^{ij} \approx 270$~ns. Here we use 50~nm film with a lateral dimension of 52~$\mu$m (length) $\times$ 47$\mu$m (width) $\times$ 50~nm (thickness), the equilibrium time between the heater and the thermometer will be even smaller. The estimated thermal equilibrium time is much shorter than the measured time constants. Thus, the equilibration between the heater, the thermometer and the pad cannot explain the appearance of additional time constants.

\section{Effect of heating pulse amplitude on thermal relaxation}
\label{appendix:c}
\begin{figure}[ht]
\includegraphics[width=\columnwidth]{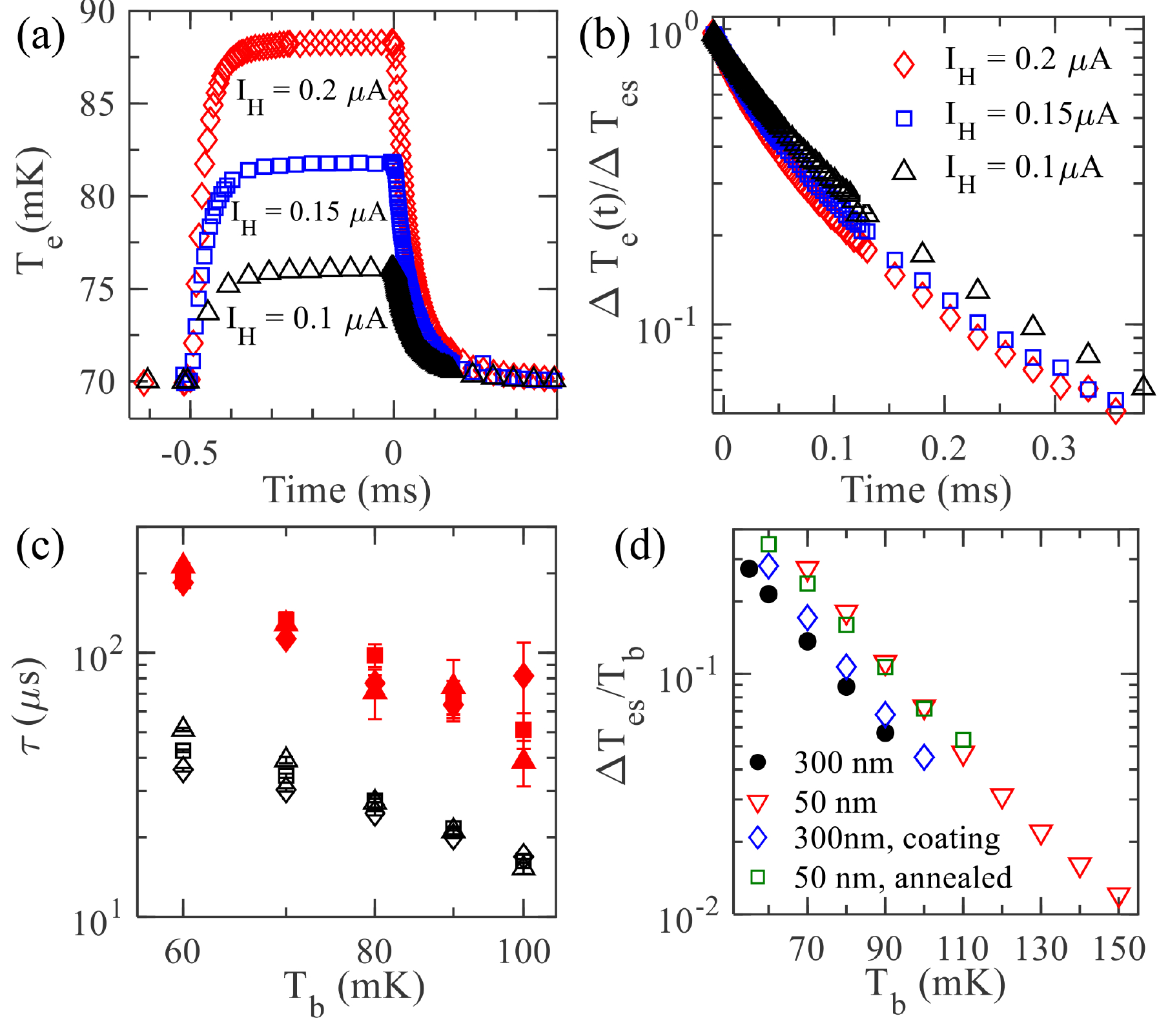}
\caption{Impact of heat magnitude on thermal relaxation. (a), (b) Electron temperature and normalized electron temperature increases $\Delta T_e(t)/\Delta T_{es}$ as a function of time interval between probing current pulse $I_p$ and heating current pulse $I_H$. The heating current pulse $I_H$ are of 0.1 (triangles), 0.15~$\mu$A (squares), 0.2~$\mu$A (diamonds). (c) Temperature dependence of the derived two thermal relaxation times $\tau_L$ (solids) and $\tau_S$ (hollows) for heating current $I_H$ of 0.1~$\mu A$ (triangles), 0.15~$\mu A$ (squares) and 0.2$\mu A$ (diamonds). (d) Ratio of the steady-state temperature increment $\Delta T_{es}$ to bath temperature $T_b$ at varying $T_b$.}\label{Fig7}
\end{figure}
In the experiment, we have varied the magnitude of the heating pulses. Figure.~\ref{Fig7} (a) shows the measured electron temperature with $I_H$ varying from 0.1 to 0.2 $\mu$A at $T_b$ = 70~mK. The thickness of the Cu film in this measurement is 300~nm and its top surface has been coated with 5~nm of Al. The steady-state temperature increment $\Delta T_{es} = T_{es} - T_b$ grows from 6 to 18 ~mK with $I_H$ increasing from 0.1 to 0.2 $\mu$A. The time dependence of the normalized temperature change $\Delta T_{e}(t)/\Delta T_{es}$ is shown in Fig.~\ref{Fig7} (b). Fitting the relaxation process with Eq.~\ref{eq1}, we derive the two thermal relaxation times $\tau_S$ and $\tau_L$ and plot them as a function of temperature in Fig.~\ref{Fig7} (c). To derive Eq.~\ref{eq1} we have assumed the temperature increase $\Delta T_e$ is much smaller than $T_e$. If high heating power is applied, this assumption is no longer valid, and, for example, the time $\tau_S$ extracted from the fits becomes smaller as it decreases at higher temperatures (see Fig.~\ref{Fig7} (c)). However, the long relaxation time $\tau_L$ is almost not affected by the heating magnitude. Figure.~\ref{Fig7} (d) shows the ratio of steady state temperature increment $\Delta T_{es}$ to the bath temperature $T_b$. In most of case, $\Delta T_{es}$ is more than one order of magnitude smaller than $T_b$. Thus, we can neglect the influence of heating amplitude on the relaxation times.

\section{Effect of annealing on thermal relaxation}
\label{appendix:d}
\begin{figure}[h]
\includegraphics[width=\columnwidth]{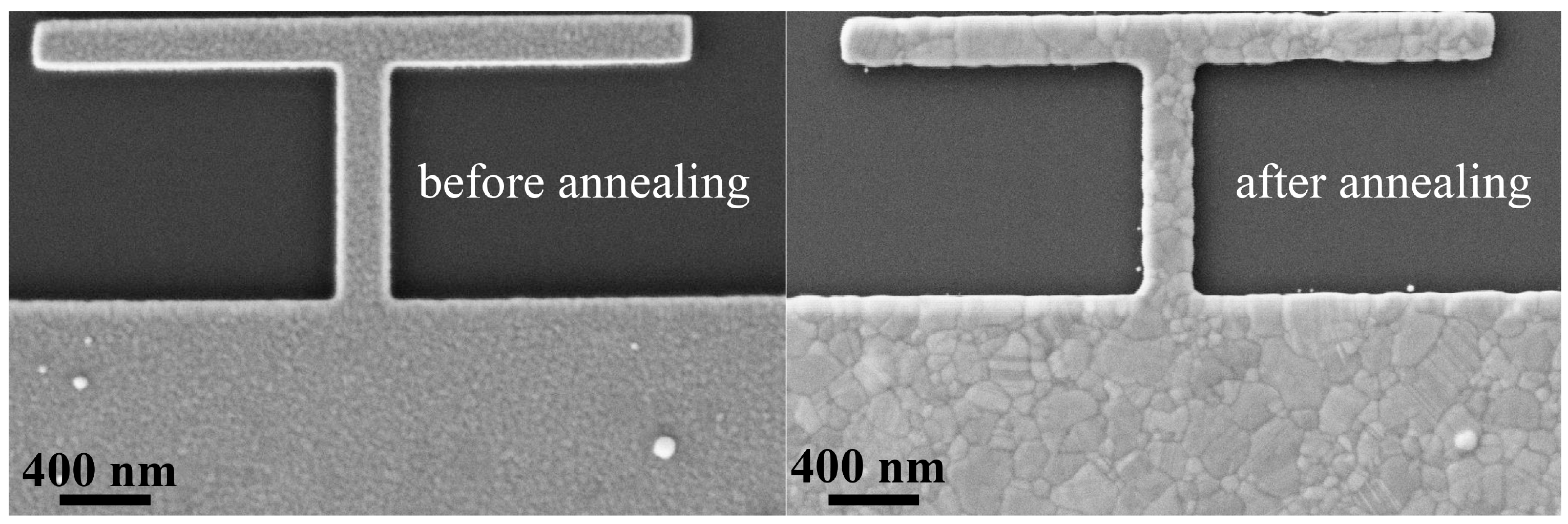}
\caption{SEM images show the grain growth after film annealing. \label{Fig8}}
\end{figure}

We have measured a 50~nm thick annealed Cu film to study the effect of grain size and grain interface on the dynamic thermal relaxation. The annealing of the film was carried out in \chem{N_2} ambient at 350\textdegree{}C for 20 mins. Clean Al contacts to the Cu film have been fabricated after the annealing. SEM images of the Cu film before and after annealing are shown in Fig.~\ref{Fig8}. Growth of the grain sizes after annealing is evident.

\begin{figure}[h]
\includegraphics[width=\columnwidth]{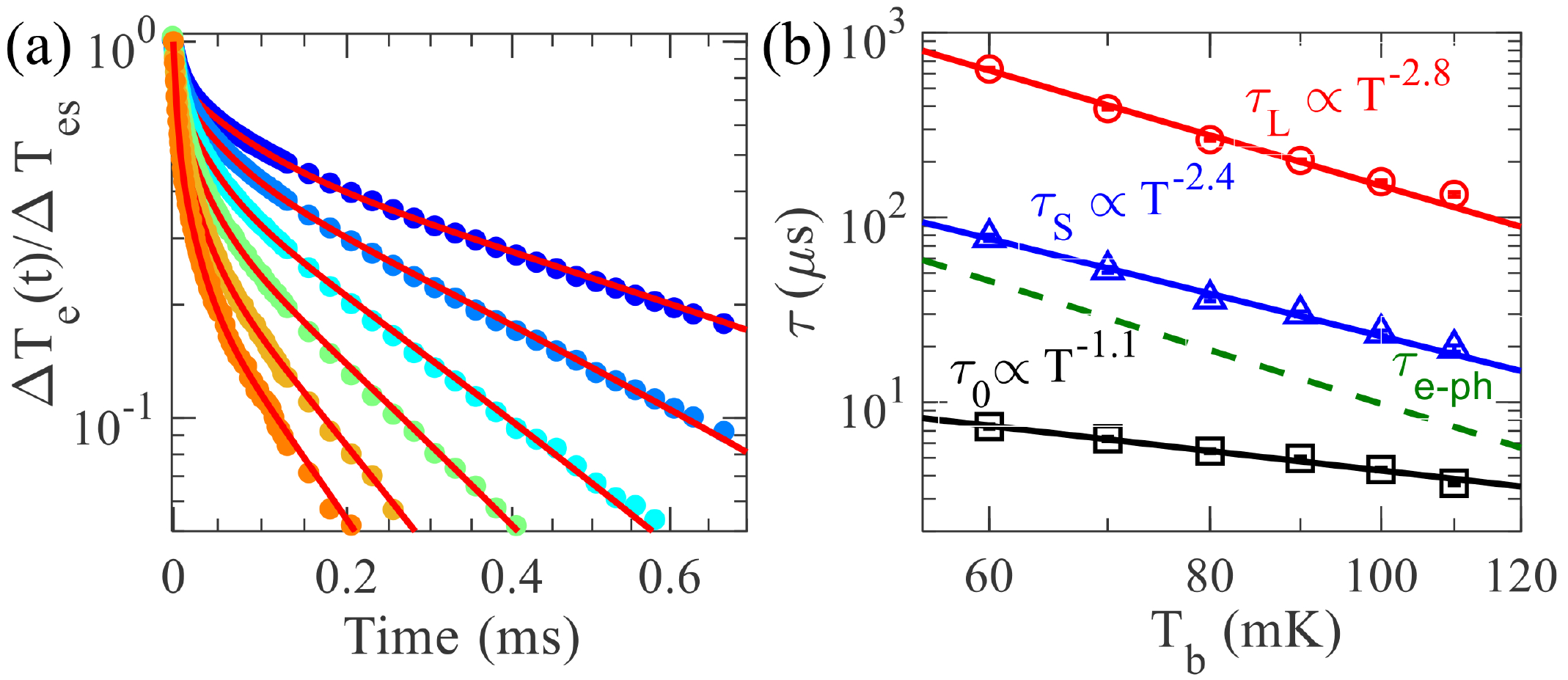}
\caption{Measurement results of the annealed 50~nm Cu film. (a) $\Delta T_e/\Delta T_{es}$ as a function of time. $T_b$ varies from 60~mK (up-blue) to 120~mK (bottom-red) with 10~mK temperature interval. Red lines are fits with Eq.~\ref{eqD1}. (b) Temperature dependence of the three thermal relaxation times. Solid lines are the fits to their temperature dependence. Dashed line is the calculated e-ph thermal relaxation time with equation $\tau_{e-ph} = \gamma/5\Sigma T^{-3}$, Sommerfeld constant $\gamma_{Cu}$ = 98~JK$^{-2}$m$^{-3}$,  electron-phonon thermal coupling constant $\Sigma_{Cu}$ = 2~nWK$^{-5}$m$^{-3}$.}\label{Fig9}
\end{figure}

Interestingly, we have also found that the thermal relaxation process is significantly affected by annealing. Figure~\ref{Fig9} shows the experimental results for the annealed 50~nm Cu film. We find that the thermal relaxation process cannot be well fitted with the double exponential decays defined in Eq.~\ref{eq1}. Instead, one needs three exponents in order to fit the data. In Fig.~\ref{Fig9} (a), we plot the $\Delta T_e(t)/\Delta T_{es}$  for the annealed film together with  fits with the expression

\eq{
	\Delta T_e (t)/\Delta T_{es} = ae^{-t/\tau_L} + be^{-t/\tau_S} + (1-a-b)e^{-t/\tau_0},
 \label{eqD1}
}
where $a$, $b$ are constants. The extracted three time constants $\tau_L$, $\tau_S$ and $\tau_0$ are plotted in Fig.~\ref{Fig9} (b). We find that the longest two time constants ($\tau_L$, $\tau_S$ ) are on the same order of the two time constants in the films not subject to annealing. The shortest time $\tau_0$ is about 6 - 10 times smaller than $\tau_{S}$. Measurement results for the annealed film cannot be explained by the introduced phenomenological thermal model (\ref{relaxation}). However, the experimental results clearly indicate the importance of the grain structure of the film for thermal relaxation. 

\section{Double exponential thermal relaxation without additional thermal reservoir}
\label{appendix:e}
Previous steady-state measurements have shown that for metallic films evaporated on a silicon substrate the thermal conductance between electrons and phonons, $G_{e-ph}$, could be comparable with the thermal boundary conductance between phonons in the film and the substrate, $G_{ph-ph}$ \cite{Wang2019}. One can argue that in this case, two different thermal relaxation times can arise already from the dynamics 
of the coupled electron and phonon subsystems without the additional thermal reservoir. In order to test this hypothesis, we consider the following equations  
describing the thermal balance in the film
\begin{eqnarray}
C_{ph}\frac{dT_{ph}}{dt} &=& G_{e-ph}(T_e - T_{ph}) - G_{ph-ph}(T_{ph} - T_{b}),
\nonumber\\
C_e\frac{dT_e}{dt} &=& -G_{e-ph}(T_e - T_{ph}) + P_H(t).
\label{eqE1}
\end{eqnarray}
Solving Eqs.~\ref{eqE1} we obtain the result similar to Eq. (\ref{eq1}), 
\eq{
\Delta T_e (t)/\Delta T_{es}= ce^{-\frac{t}{\tau_1}} + (1-c)e^{-\frac{t}{\tau_2}}, \label{eqE2}
}
where $\Delta T_e(t) = T_e(t) - T_b$ is the time dependent electron temperature increase, and $c$ is a constat. 
The two time constants $\tau_1$ and $\tau_2$ read
\begin{eqnarray}
	\tau_1 &\approx& \frac{C_e}{G_{e-ph}}+\frac{C_e}{G_{ph-ph}}. \label{eqE3}
\\
	\tau_2 &\approx& \frac{C_{ph}}{G_{e-ph} + G_{ph-ph}} . \label{eqE4}
\end{eqnarray}
The phonon heat capacity can be estimated as 
\begin{equation}
C_{ph} = 2\pi^2k_B^4T^3/(5\hbar^3c^3),
\label{Cph}
\end{equation} 
where $k_B$ is the Boltzmann constant, $\hbar$ is the reduced Planck constant, and $c$ is the speed of sound in the film. In the temperature range from 50 to 200~mK, the electronic heat capacity $C_e$ is 4000 to 250 times larger than the phonon heat capacity $C_p$. Hence, $\tau_2$ is two to three orders of magnitude of smaller than $\tau_1$, and it cannot be resolved with our thermometer. The ratio of the two experimentally observed time constants, $\tau_L/\tau_S$, roughly remains within one order of magnitude. In Fig. \ref{Fig4}a we plot $C_{ph}$ together with the heat capacity of the reservoir extracted from the data $C_d$, it is clear that $C_{ph}\ll C_d$. Thus, we conclude that the simple model (\ref{eqE1}) cannot explain our observations due to the very small value of the phonon heat capacity $C_{ph}$.

\section{Pre-factor $a$ extracted from the fits}
\label{appendix:f}
\begin{figure}[h]
\includegraphics[width=0.65\columnwidth]{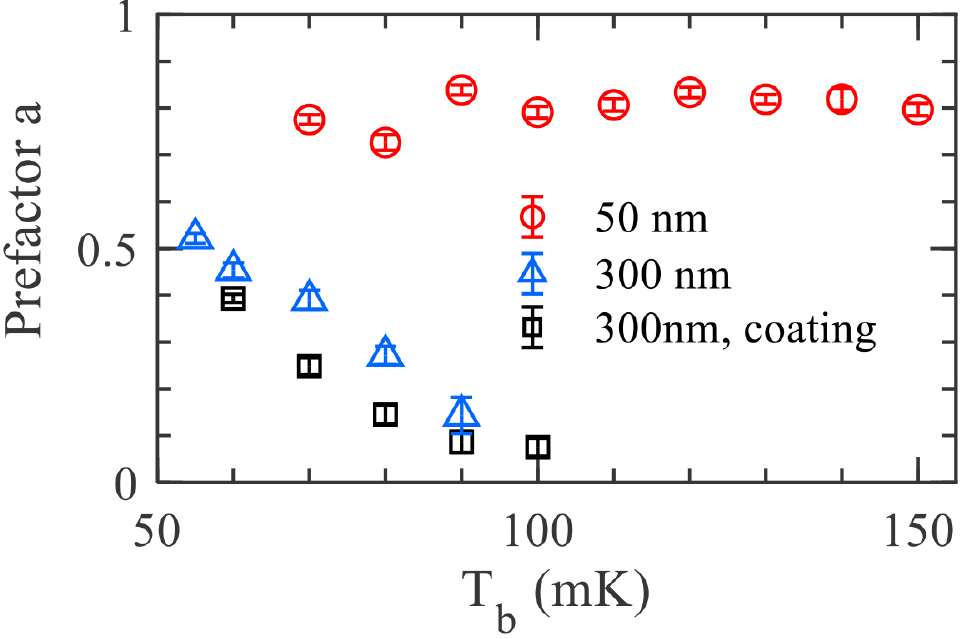}
\caption{Temperature dependence of the prefactor $a$ derived from the fits to experiment data with Eq.~\ref{eq1}.}\label{Fig10}
\end{figure}

We have introduced the thermal model considering film electrons thermally coupled to the phonons and an additional thermal reservoir. Within this model, the dynamic relaxation process of Cu films is described by Eq.~\ref{eq1}. Fitting the experimental data with Eq.~\ref{eq1}, we obtain the two time constants $\tau_L$ and $\tau_s$, as well as the prefactor $a$ in front of the slowly decaying exponent. In Fig.~\ref{Fig10}, we show its temperature dependence obtained for the two 300~nm Cu films and the as-deposited 50~nm Cu film. For the annealed 50~nm Cu film, one needs to used three exponents to fit the data, as shown in the previous section. We find that for the 50~nm Cu film $a$ remains close to 0.8 without obvious temperature dependence. However, for the two 300~nm Cu films $a$ drops from 0.5 at 55~mK to close to 0 at 100~mK. The origin of such temperature dependence remains unclear and requires further investigation. The high value of $a$ observed in 50~nm Cu film suggests that the dominant texture is this film was $\langle 110\rangle$. In contrast, in the two 300~nm Cu films with $a<0.5$, the $\langle 111\rangle$ texture probably dominates.

\bibliography{relaxation2019}
\end{document}